**Title:** MEMS-actuated Metasurface Alvarez Lens

**Author List:**

*Zheyi Han*[1,4], email: zh25@uw.edu

*Shane Colburn*[1], email: scolbur2@uw.edu

*Arka Majumdar*[1,2], email: arka@uw.edu

*Karl F. Böhringer*[1,3,4,*], email: karlb@uw.edu, phone: 206-221-5177, fax: 206-543-3842

[1]Department of Electrical and Computer Engineering, University of Washington, Seattle, Washington 98195, USA.

[2]Department of Physics, University of Washington, Seattle, Washington 98195, USA.

[3]Department of Bioengineering, University of Washington, Seattle, Washington 98195, USA.

[4]Institute for Nano-engineered Systems, University of Washington, Washington 98195, USA.

[*]Correspondence to: Karl F. Böhringer

**Abstract**

Miniature lenses with a tunable focus are essential components for many modern applications involving compact optical systems. While several tunable lenses have been reported with various tuning mechanisms, they often face challenges with respect to power consumption, tuning speed, fabrication cost, or production scalability. In this work, we have adapted the mechanism of an Alvarez lens – a varifocal composite lens in which lateral shifts of two optical elements with cubic phase surfaces give rise to a change in the optical power – to construct a miniature, microelectromechanical system (MEMS)-actuated metasurface Alvarez lens. Implementation based on an electrostatic MEMS generates fast and controllable actuation with low power consumption. The utilization of metasurfaces – ultrathin and subwavelength-patterned diffractive optics – as optical elements greatly reduces the device volume compared to systems using conventional freeform lenses. The entire MEMS Alvarez metalens is fully compatible with modern semiconductor fabrication technologies, granting it the potential to be mass-produced at a low unit cost. In the reported prototype operating at 1550 nm wavelength, a total uniaxial displacement of 6.3 µm was achieved in the Alvarez metalens with a direct-current (DC) voltage application up to 20 V, which modulated the focal position within a total tuning range of 68 µm, producing more than an order of magnitude change in the focal length and a 1460-diopter change in the optical power. The MEMS Alvarez metalens has a robust design that can potentially generate a much larger tuning range without substantially increasing the device volume or energy consumption, making it desirable for a wide range of imaging and display applications.

**Introduction**

Miniature, focus-tunable optical elements are crucial for various applications where the size and weight of the optics are highly constrained. For example, the optical components used in mobile applications, entertainment, robotics and surveillance, aerospace, and biomedical systems often need to be tunable within a large focal length range while maintaining reliability and cost-effectiveness[1,2]. Conventional tunable imaging systems often have either bulky optical elements that are challenging to mass produce, such as liquid tunable lenses[3], or tuning mechanisms that require a large volume for the actuation of elements, such as physically translating the lens along the optical axis[4]. A miniaturized optical element together with an efficient tuning mechanism is essential for designing a compact tunable lens system.

Metasurface optics are novel replacements for bulky freeform optical elements with a greatly reduced volume and high compatibility with modern microfabrication processes. These elements are 2D, quasi-periodic arrays of subwavelength scatterers arranged to arbitrarily control the phase, amplitude, and polarization of electromagnetic waves by imparting an abrupt, spatially varying phase profile on the incident light[5]. With metasurfaces, complex and bulky geometric lens curvatures can be converted into a discretized spatial phase profile on a flat, wavelength-scale thick surface[6-10]. The corresponding fabrication process can employ mature semiconductor nanofabrication technologies that are readily scalable for mass production.

There have been multiple works that have attempted to modulate the optical power of metasurfaces with various mechanisms. Ee *et al*. designed and fabricated a mechanically reconfigurable metasurface on a stretchable membrane for the visible frequency range, with the deformation induced by four clamps individually mounted on translation stages[11]. The design by She *et al*. combined metasurface optics with a dielectric elastomer actuator to create an electrically tunable focal length by electrostatically compressing the elastomer in the direction of its thickness, leading to a lateral expansion in both the elastomer and the carried metasurface; however, due to the limitations in the elastomer viscoelasticity, charge transfer, and dissipation time in the electrodes, the response time of their design was capped at 33 ms[12]. Roy *et al*. designed a microscale tunable mirror by placing a reflective metasurface lens on top of a 2D electrically actuated scanner platform[13]. The angular rotation of the mirror plate scans the focal position of the metasurface lens by several degrees in a 3D space. The manual assembly required to integrate the individual micromirror and the external microelectromechanical system (MEMS) platform inevitably increases the overall fabrication time, rendering the device less desirable for mass production. Arbabi *et al*. fabricated and demonstrated a tunable metasurface doublet lens realized by implementing vertical MEMS actuation. Their system comprised both converging and diverging metasurface lenses assembled on top of each other; this system can provide a tunable focal distance that is controlled by the electrostatically varied separation between the two lenses[14]. While this method can provide a change in the focal length that is inversely proportional to the vertical displacement, the fabrication entails an initial separation between the lenses to reserve space for the designed actuation, which limits the overall tuning range if a thin device is desired. Additionally, the parallel plate actuator in their design can only increase the focal length from the initial value by decreasing the optics separation rather than tuning in both directions.

A separate approach for creating a miniature lens system with tunable optical power is based on Alvarez lenses[15-17]. A typical Alvarez lens contains a pair of optical elements with complementary cubic surface profiles. Optical power modulation is achieved by the relative lateral displacement between the two optical elements in the direction perpendicular to the optical axis. In this way, the focal length change is inversely proportional to the lateral displacement, and the focal length can be tuned in both directions. Previously, some researchers integrated a pair of miniature, freeform optics with cubic surface profiles with a MEMS actuator to electrostatically displace the optical elements laterally for focus tuning within a compact package[18,19]. In these works, however, the lens elements and the MEMS structure were manufactured separately and then manually assembled, requiring additional time and cost. Moreover, the conventional refractive optics used in these Alvarez systems require complex 3D geometries to produce the desired cubic surfaces for optimal performance and mounting structures for assembly. This complexity often requires complicated fabrication and at the same time can be a substantial obstacle for further miniaturization.

A possible solution to create a miniature, focus-tunable optical system compatible with large-scale microfabrication is to convert the 3D Alvarez optical elements into a pair of 2D metasurface optics with complementary cubic surface profiles; these optics can be integrated with a microscale MEMS-actuated structure to introduce lateral displacement for optical modulation. Metasurface Alvarez lenses have shown great potential for substantial focal length control with small relative lateral shifts. Recent studies have demonstrated an Alvarez metalens inducing a 2.5-mm focal tuning range in visible wavelengths over a 100-µm actuation range with a 150-µm wide square aperture[1] and another inducing a 6.62-cm focal tuning range for infrared (IR) wavelengths over a 2.75-mm actuation range with a 1-cm wide square aperture[2]. Currently, however, displacements have been performed using optomechanical stages that are actuated manually, while a compact and electrically controllable solution has not been demonstrated.

In this work, we present a silicon nitride metasurface-based Alvarez lens actuated by an electrostatic MEMS platform capable of producing a focal length modulation that is 10 times larger than the actuated displacement. The metasurfaces are fabricated with high-throughput stepper lithography, and the entire structure is fully compatible with well-developed semiconductor nanofabrication technologies, providing high reliability, low cost, and the potential for production scalability.

**Results**

Design of the MEMS-actuated Metasurface Platform

A miniature MEMS electrostatic actuator is implemented to laterally displace one of the complementary Alvarez metasurface optics relative to the other with high precision and speed. Since a conventional Alvarez lens requires the symmetric actuation of both phase masks, in our configuration, where only one of the Alvarez plate pairs is actuated to enhance robustness during prototype fabrication and operation, a lateral shift of the optical axis, albeit a small one, is also induced during focal length modulation. The actuator design for the moving metasurface is based on conventional linear comb drives[20] and exact constraint folded flexures[21] to introduce the electrostatic driving force and to generate the mechanical restoring force, respectively.

Electrostatic comb drives can induce an efficient, controllable displacement at low power; hence, they are often popular candidates for microscale devices designed for sensing and micropositioning. A typical design of a comb drive actuator consists of two comb structures with interdigitated fingers, of which one is fixed while the other is connected to a compliant suspension. A voltage applied across the two comb structures will lead to a deflection of the movable comb drawn by the electrostatic forces created in the capacitors between the fingers with different potentials. Since the electrostatic forces are created by capacitive interactions without any physical contact between the two combs, there is no direct current (DC) flowing in the device, realizing low-power actuation. The design of using an array of small fingers instead of one single large parallel plate capacitor significantly increases the area for capacitive interaction and hence the resultant electrostatic force with a given applied voltage. Moreover, the linear comb design allows controlled motion over a larger displacement range compared to the parallel plate capacitor. The substrate and the mobile comb are usually grounded to prevent the development of electrostatic pull-down forces acting on the structures in suspension, and only the stationary comb is at a nonzero potential to generate an electrostatic driving force between the combs.

With the assumption that the fringing field effect is negligible, the total capacitance between the interdigitated fingers in each comb drive can be estimated as

$$C_{total} = 2N \frac{\varepsilon l h}{d}, \tag{1}$$

where $N$ is the number of finger pairs and $\varepsilon$ is the permittivity of the medium between the fingers that have an interdigitating length $l$, a height $h$, and a separation $d$. With a voltage $V$ applied across the comb drive, the corresponding electrostatic force becomes

$$F_{el}(V) = \frac{dW(V)}{dl} = \frac{d}{dl}\left(\frac{1}{2}C_{total}V^2\right) = \frac{N\varepsilon h V^2}{d}. \tag{2}$$

If the movable comb, carrying the metasurface platform, is connected to some mechanical flexure with a spring constant $k$, then at equilibrium, the lateral displacement of the movable comb is

$$x(V) = \frac{F_{el}(V)}{k} = \frac{N\varepsilon h V^2}{kd}. \tag{3}$$

Note that the actuated displacement is linearly proportional to the electrostatic force, making comb drives better candidates for large displacements than parallel plate capacitors. Since the physical dimensions and material properties do not change after fabrication, with proper device calibration, the exact actuated displacement can be predicted by Eq. (3) at any given actuation voltage. Such controllability makes comb-drive actuators very attractive structures for applications involving micropositioning.

For unidirectional actuation, we need a suspension platform, which is compliant along the direction of the desired movement but stiff in orthogonal directions. To enable this compliance, we employ the mechanical structures of the exact constraint folded flexure[21]. By constraining the displacement of the intermediate body with a 1:2 lever, we can reduce the development of extensional axial forces, preventing sideway motion or rotation when a lateral displacement is actuated. The mechanical stiffness of the actuator in the direction of the lateral displacement can be estimated with the lateral spring constant of the folded flexure[22]

$$k_x = \frac{2Ehb^3}{L^3}, \qquad (4)$$

where $E$ is Young's modulus of single-crystal silicon, $h$ is the device layer thickness (also the spring beam and finger height), $b$ is the width of spring beams, and $L$ is the length of individual beams. For all our actuators, $L = 500$ μm, $b = 4$ μm, and $h = 5$ or $6$ μm are the design parameters. With $E = 130$ GPa in the silicon-100 plane[23], an actuator will have a lateral stiffness of 0.666 N/m (0.799 N/m) for an $h$ of 5 μm (6 μm). The effect of the additional 1:2 lever is difficult to calculate analytically but is expected to slightly increase the stiffness value. We performed an eigenfrequency analysis on our fundamental MEMS actuator design using COMSOL Multiphysics, which predicted a mechanical resonance at 3.4 kHz, showing potential for high-speed, high-precision varifocal tuning in sensing and display applications.

Figure 1 shows the scanning electron microscope (SEM) and optical images of a fabricated MEMS actuator carrying the mobile metasurface of the Alvarez pair, which is then aligned and bonded on top of the complementary metasurface fabricated on a stationary substrate. The microfabrication flow can be found in the Materials and Methods section. As shown in Figure 1a, the actuator design used here has two sets of comb drives and folded flexures on the opposite sides of the platform to double the actuation range under signals applied alternatively to both sides of the actuator and to prevent rotational displacement during the movement. Figure 1b shows a zoomed-in SEM image of a comb drive. The suspended left comb is attached to the central platform for displacement, while the stationary right comb is anchored to the substrate. One of the silicon nitride Alvarez metasurfaces is carried by the central platform attached to the suspended combs of each set of comb drives, while the other is fabricated on a stationary silicon substrate. Figure 1c shows the details of one metasurface composed of a 2D array of cylindrical silicon nitride nanoposts, generating the desired phase profile for the Alvarez optics. Figure 1d and 1e show optical images of the two Alvarez metasurfaces with complementary surface profiles. The two chips carrying the metasurfaces are overlaid on top of each other in the bonded lens system.

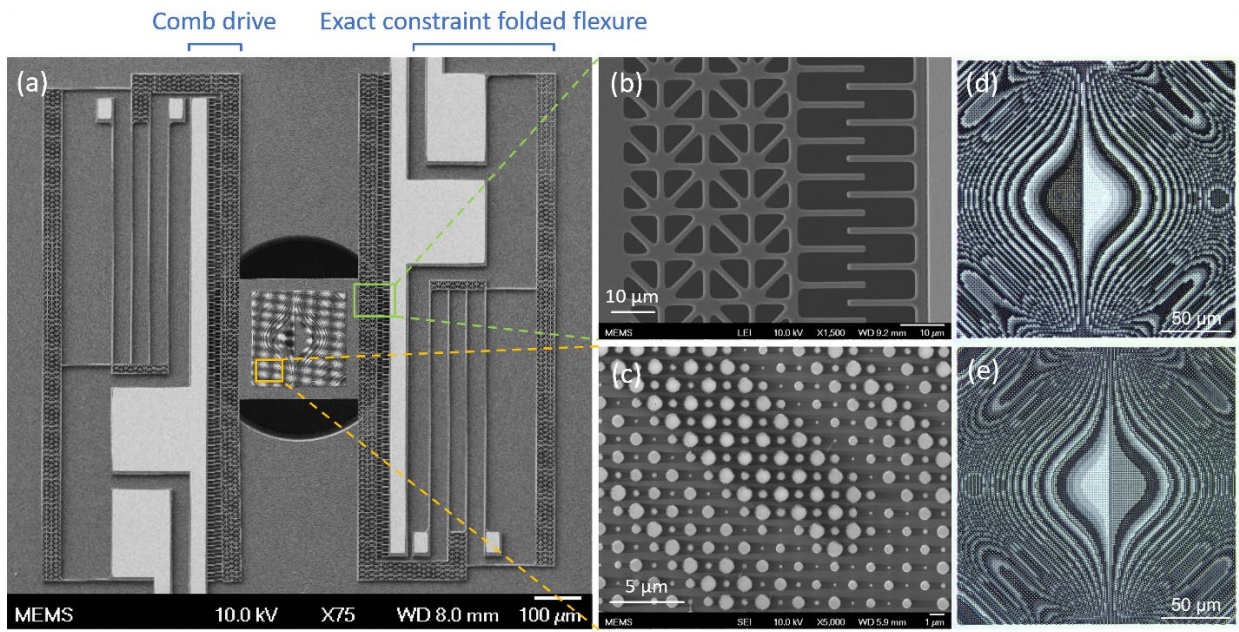

*Figure 1.* SEM image of (a) a MEMS-actuated platform carrying the movable metasurface in an Alvarez metalens, with (b) and (c) being the zoomed-in SEM images portraying the electrostatic comb drive and the metasurface nanoposts, respectively. (d, e) Optical images of the Alvarez metasurfaces with complementary cubic phase profiles that are overlaid on top of each other in the assembled Alvarez metalens and in summation impart a quadratic phase change on the incident light.

MEMS-actuated Metasurface Singlet Lens

Prior to testing the metasurface Alvarez lens, we fabricated a separate structure with a singlet converging metasurface lens with a diameter of 200 μm on the central platform of the MEMS actuator to aid motion tracking by examining the movement of the focal spot (Figure 2a). The metasurface design followed the usual procedure[24], which is outlined in the Materials and Methods section. The left and right sides of the platform were attached to comb drives for actuation. The prototype actuator was fabricated with 2 comb drive sets per side (with 125 finger pairs in each) and a device layer thickness of 6 μm. With the MEMS actuator chip mounted on a glass stage, a superluminescent diode (SLD) source of 1550 nm illuminated the metasurface lens from the back. The resulting focal spot was monitored by an IR camera in front as an indication of the lens location (the experimental setup is described in the Materials and Methods section). Since the focal spot was located approximately 200 μm in front of the metasurface, for demonstration of the relative lateral position of the metasurface and the focal spot, Figure 2a shows an image overlay of the device surface plane and the focal plane. A DC voltage ramping from 0 V to 18 V and then back to 0 V with a step size of 500 mV was applied to the right comb drive sets of the actuator to electrostatically pull the central platform. The maximum testing voltage was chosen based on the actuator dimensions to avoid electrostatic pull-in between the combs. The increasing voltage pulled the metasurface and the focal spot toward the right. By turning down the voltage, the focal spot relaxed back to its neutral position equilibrated by the spring force in the flexures. Figure 2b plots both the displacement of the focal spot and the applied voltage as time elapses, showing a clear positive correlation between the actuated displacement and the actuation voltage. The IR camera resolution (approximately 0.9 pixels per micron) limits the number of pixels available for focal spot tracking, giving rise to the distinctive steps in the focal location analysis. The symmetric displacement during voltage ramp-up and ramp-down in Figure 2b shows the absence of hysteresis and verifies the reversibility of the platform actuation. The linear fit of the focal displacement (averaged over all captured video frames at each applied voltage) as a function of the voltage squared (Figure 2c) further confirms that the actuated displacement is quadratically dependent on the voltage applied. The stiffness of the platform actuator can be calculated from the slope in Figure 2c according to Eq. (3), giving a spring constant of 0.713 N/m for this electrostatic actuator, which is 11% smaller than the analytically calculated value presented earlier. This result indicates a softer device than predicted, probably due to fabrication discrepancies resulting in thinner structures. While we observe a displacement of 3.1 μm under a voltage application of 18 V, if a voltage is applied to the left rather than the right side of the actuator, the metasurface can be displaced in the opposite direction, doubling the range of the physical motion.

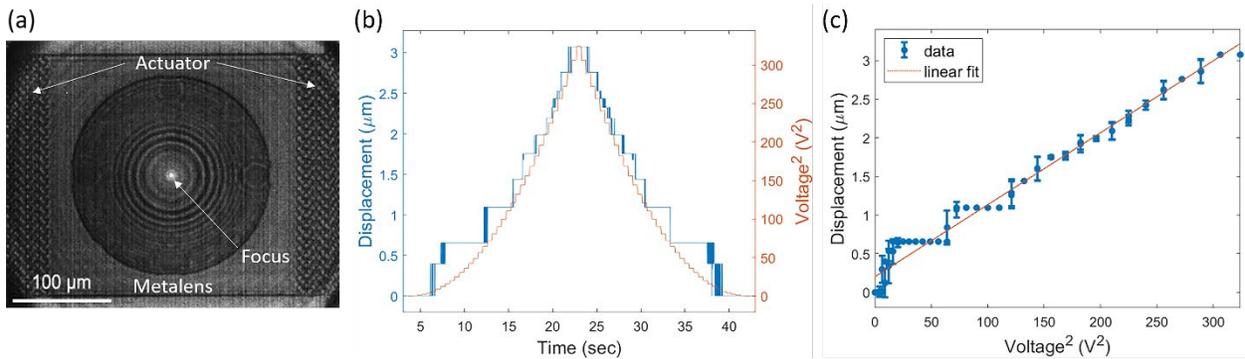

*Figure 2. (a) Overlaid IR image showing both the device and focal planes of a singlet metalens on the MEMS-actuated platform. (b) Actuated focal displacement and applied voltage over time. The general trend of the actuated displacement follows the square of the applied voltage. (c) Average actuated displacement as a function of applied voltage squared during both loading and unloading, showing a linear dependence. No hysteresis behavior is observed with voltage loading and unloading, verifying the reversibility of the actuation process.*

Design of the Metasurface Alvarez Lens

With the electromechanical performance of the MEMS-actuated platform verified using a singlet metalens, we designed and fabricated the Alvarez lens device integrated into the MEMS platform consisting of two separate IR metasurfaces with apertures of 200 μm. When positioned with their surfaces parallel and aligned along the optical axis, the two Alvarez metasurfaces in conjunction form a singlet metalens that imparts a quadratic phase profile onto the incident wavefront, as shown in Figure 3a. The phase profiles of the regular and inverse metasurfaces of the Alvarez pair have the expressions[2]

$$\varphi_{reg}(x,y) = -\varphi_{inv}(x,y) = A\left(\frac{1}{3}x^3 + xy^2\right). \tag{5}$$

A lateral displacement between the two metasurfaces modulates the optical power of the summed quadratic lens, where the constant $A$ denotes the cubic phase strength chosen to produce a desirable focal length tuning range for a given aperture size and a designed displacement range. In a typical Alvarez lens, the phase masks are simultaneously translated in opposite directions by a displacement $d$ (hence creating a total center-to-center offset of $2d$ between the phase masks), producing a quadratic lens in summation with the expression

$$\varphi_{Alvarez}(x,y) = \varphi_{reg}(x+d,y) + \varphi_{inv}(x-d,y) = 2Ad(x^2+y^2) + \frac{2}{3}Ad^3. \tag{6}$$

Neglecting the constant $d^3$ term, we can relate the summed quadratic phase to that of a spherical singlet lens and express the focal length of the Alvarez lens as a function of the symmetric lateral displacement at a designed operating wavelength $\lambda$ as

$$f(d) = \frac{\pi}{2\lambda A d}. \tag{7}$$

A zero summed phase corresponding to a flat lens will be produced if the centers of the two Alvarez phase patterns are exactly aligned. For the specific Alvarez lens fabrication and experiments presented in this work, to accommodate data collection over our achievable actuation range with the current actuator designs, an initial offset of $d = 10$ μm is introduced to each of the two metasurfaces in opposite directions, hence giving a total of 20 μm initial center-to-center lateral offset between the Alvarez phase masks, as captured in Figure 3a.

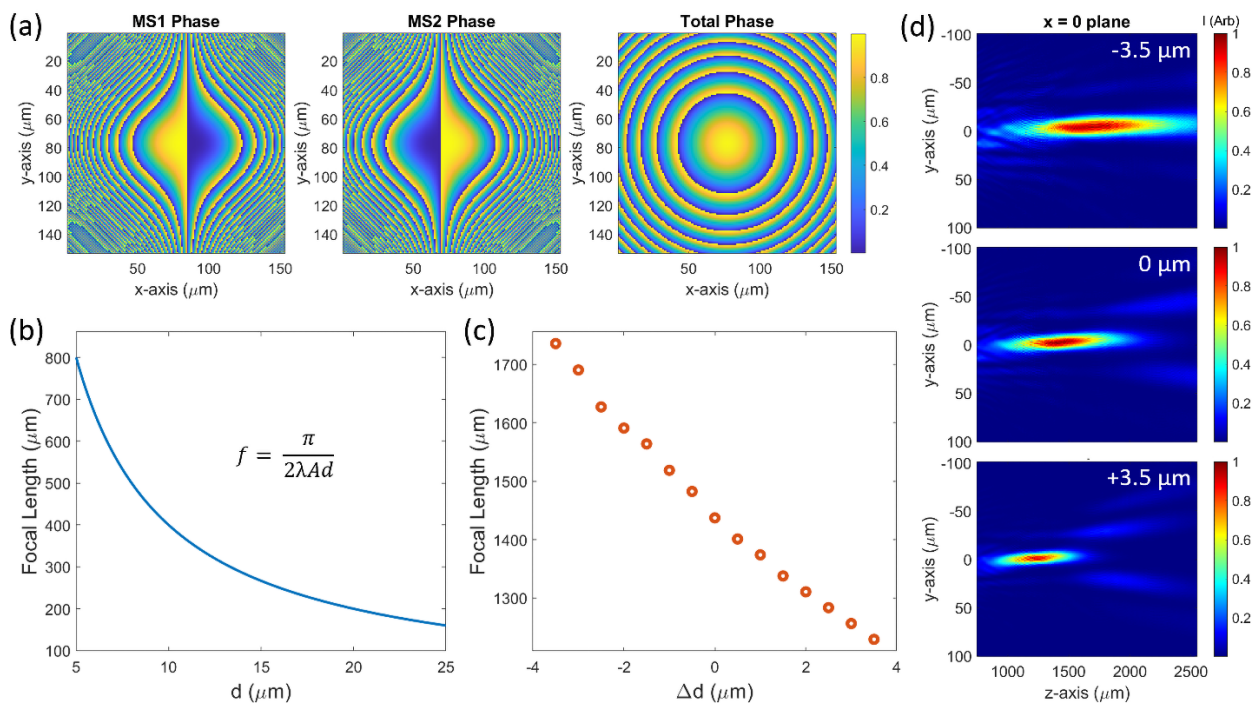

***Figure 3.*** *Metasurface Alvarez lens design and simulation. (a) Two 200 μm × 200 μm Alvarez metasurfaces with complementary cubic phase profiles and their summed quadratic phase when overlaid on top of each other. The phase values are normalized to 2π. (b) Ideal Alvarez focal tuning behavior as a function of the lateral symmetric displacement d (giving a total center-to-center offset of 2d) between the optical elements theorized with a negligible axial separation gap, an operating wavelength λ = 1550 nm and a cubic parameter A = 2.5335 × $10^{14}$ $m^{-3}$. (c) Simulated focal tuning behavior when a small lateral displacement is introduced in addition to the initial center-to-center offset of 20 μm, assuming a 50 μm axial separation gap between the two metasurfaces from fabrication. The focal position is calculated from the weighted centroid of the high-intensity regions. (d) Simulated change in the focal location and profile when the two metasurfaces are laterally displaced at –3.5 μm, 0 μm, and +3.5 μm.*

Figure 3b illustrates the ideal tuning behavior of a typical Alvarez lens assuming no axial separation exists between the two complementary phase masks[2], showing a nonlinear dependence of the focal position on the lateral offset. A larger center-to-center lateral offset between the Alvarez phase masks gives rise to a more rapidly varying phase profile, corresponding to a lens with a shorter focal length[1]. Figure 3c, instead of the ideal Alvarez behavior, shows the simulated behavior of the designed metasurface Alvarez lens under a more realistic situation in which an axial gap of 50 μm is introduced between two metasurfaces from the bonding process and the lateral displacement occurs in small microscale steps, as expected from the electrostatic MEMS actuator with voltage ramping in steps. Note that in our MEMS-actuated Alvarez metalens, to increase device robustness, only one metasurface phase mask is actuated relative to the other, instead of both phase masks being actuated simultaneously in opposite directions as in a conventional Alvarez lens. Hence, the actuated displacement $\Delta d$ presented in Figure 3c adds to the total center-to-center offset between the two metasurface phase plates, accompanied by a small translational shift of the optical axis during tuning. The 50 μm axial separation gives rise to diffraction in the space between two metasurfaces, leading to the distortion and shifting of the focal spot along the optical axis, as illustrated in Figure 3d. Specifically, the separation between two metasurfaces deviates the summed effects of the Alvarez elements from a simple phase addition. Instead of focusing on a single point, as expected by a singlet lens, the light incident on the Alvarez stack creates multiple closely spaced intensity maxima, forming an elongated and aberrated intensity cluster along the optical axis (Figure 3d). We denote the focal length by the centroid of the high-intensity clusters and plot the location of the centroid as a function of the displacement in Figure 3c. The simulation reveals close-to-linear tuning of the focal length with submicron displacement steps.

MEMS-actuated Metasurface Alvarez Lens

A full metasurface Alvarez lens was fabricated with the process flow described in the Materials and Methods section, with one Alvarez metasurface on the actuator membrane and the other on a stationary surface. The electrostatic actuator used for the Alvarez lens had 2 comb drive sets per actuation side with a device layer thickness of 5 μm. The focal tuning of the metasurface

Alvarez lens was measured with the Alvarez chip mounted on a translation stage and illuminated with a 1550 nm SLD source. A voltage between 0 and 20 V was applied to the comb drives to actuate the central metasurface platform, changing the lateral offset between two Alvarez metasurfaces. The maximum testing voltage was chosen based on the actuator dimensions to avoid electrostatic pull-in between the combs. At any point in time, we actuated the comb drives only on one side: the right-side actuation increased the lateral offset, whereas the left-side actuation decreased the lateral offset. Note that we always applied a positive DC voltage in this experiment but denoted the case of a decreasing offset (controlled by choosing which set of comb drives to actuate) between two metasurfaces by using negative voltage and displacement notations. Hence, with a negative voltage, we observe a longer focal length. Similarly, a positive voltage and displacement correspond to an increasing lateral offset and a shorter focal length. An IR camera was aligned with the Alvarez lens and the source to monitor the focal tuning behavior. The distance between the Alvarez metalens and the IR camera was controlled by a motorized linear translation stage to scan across the intensity profiles along the optical axis in microscale fine steps and to locate the focal plane tuned by the input voltage. Detailed information on the setup is included in the Materials and Methods section.

Figure 4a is a camera screenshot showing the two Alvarez metasurfaces overlaid on top of each other. The bottom metasurface sits on a stationary surface, while the top metasurface is carried by a MEMS actuator. The comb drives employed to electrostatically displace the top metasurface laterally in opposite directions are partially captured at the edges of the image. By adjusting the distance between the IR camera and the Alvarez lens, a bright focused spot, as shown in Figure 4b, was found approximately 200 μm away from the Alvarez lens. We attribute the discrepancy between the theory and experimental results on the focal tuning range of the Alvarez lens to the small aperture and the 50 μm axial separation between the metasurfaces. A side peak, as shown beside the central focal spot in Figure 4b, emerged due to the gap and misalignment between the Alvarez metasurfaces, which led to the incomplete cancellation of the cubic phase terms in the conjugate system. To test the effect of the aperture size, we measured the focus tuning of several Alvarez lenses with various aperture sizes without MEMS actuation. In these lenses, the lateral separation was actuated manually, and we found a higher tolerance to misalignment and hence a better match between the theory and experiment as the aperture size increased (see Figure S5 of the Supplementary Information). In the experiment of the MEMS-actuated Alvarez metalens, at every applied voltage, the intensity profile along the optical axis was captured over a 100 μm range around the brightest spot with an increment of 1 μm. We processed the captured images to reduce systematic noise. Baseline references at no illumination and full illumination were taken to remove anomalies due to the camera artifacts by comparison and subtraction (details included in the Supplementary Information).

With the aid of the camera screenshots capturing the actuator, as shown in Figure 4a, the actuated displacement at an applied voltage was determined from the change in the lateral position of the actuated platform relative to its neutral position using edge detection. Figure 4e and 4f plot the actuated displacement of the top metasurface at various applied voltages. The results indicate that the displacement electrostatically actuated with the comb drives is quadratically dependent on the actuation voltage, as predicted by Eq. (3), and hence can be well controlled for a calibrated system. The total stiffness of the folded flexures in the Alvarez system can be calculated from the quadratic fit of the displacement-voltage plots, giving 0.590 N/m when actuated toward the negative direction and 0.586 N/m toward the positive, verifying the symmetry of the mechanical design. The actual stiffness values are 12% smaller than the predicted values presented earlier, which we ascribe to fabrication discrepancies resulting in reasonable deviations in both electrostatic and restoring forces.

For each set of scanned data with the same actuation voltage, a 1D slice of intensities was taken from each screenshot at the same location across the central bright spot. All 1D slices were then combined into a 2D intensity plot over the scanning distance along the optical axis, as shown in Figure 4c and 4d. The slight tilting of the focal axes shown here was the result of displacing only one metasurface in the Alvarez optics pair, which introduced an additional linear term into the system's phase function, effectively acting as a beam deflector (details in the Supplementary Information). In the combined plots, a soft Gaussian filter was applied to reduce the residual stripe-patterned nonuniformity due to camera artifacts. Then, the intensities were normalized, and a threshold was taken to identify and isolate the region of high intensities. The focal position along the optical axis was calculated by locating the weighted centroid of the pixels above the defined intensity threshold to minimize the effect of background noise. Such absolute positions of the focal spots are estimations due to the difficulty in the visual determination of the zero plane at the device surface. To better visualize the focal tuning behavior, the net change in the focal position was instead plotted as a function of the actuation voltage in Figure 4g, using the corresponding neutral focal position without voltage application as the reference. Such a focal-voltage relation is then translated to a function of the actuated displacement in Figure 4h using the displacement-voltage data pairs presented in Figure 4e and 4f. Figure 4g and 4h show the focal locations identified by the weighted centroids of high-intensity clusters above the various threshold choices. A higher threshold percentile gives rise to a smaller isolated region of high-intensity pixels for the focal search; 0.1% corresponds to a bright area of 18 μm$^2$, while 0.5% corresponds to a bright area of 90 μm$^2$. As shown in Figure 4h, the Alvarez metalens produces a tunable focal length range of 68 µm with a total actuated displacement of 6.3 µm, generating a focal length tuning range that is over 10 times larger than the actuated displacement. With a neutral focal length of 216 µm, this focal tuning range corresponds to a 1460-diopter change in the optical power. At the submicron displacement steps realized with electrostatic actuation, transition states are observed in the focal tuning curves due to the axial separation between the metasurfaces from the bonding spacer. With a larger focus spot defined with a lower threshold percentile, a smoother tuning behavior can be produced with a slightly smaller range of tuning due to the asymmetric intensity distribution along the optical axis.

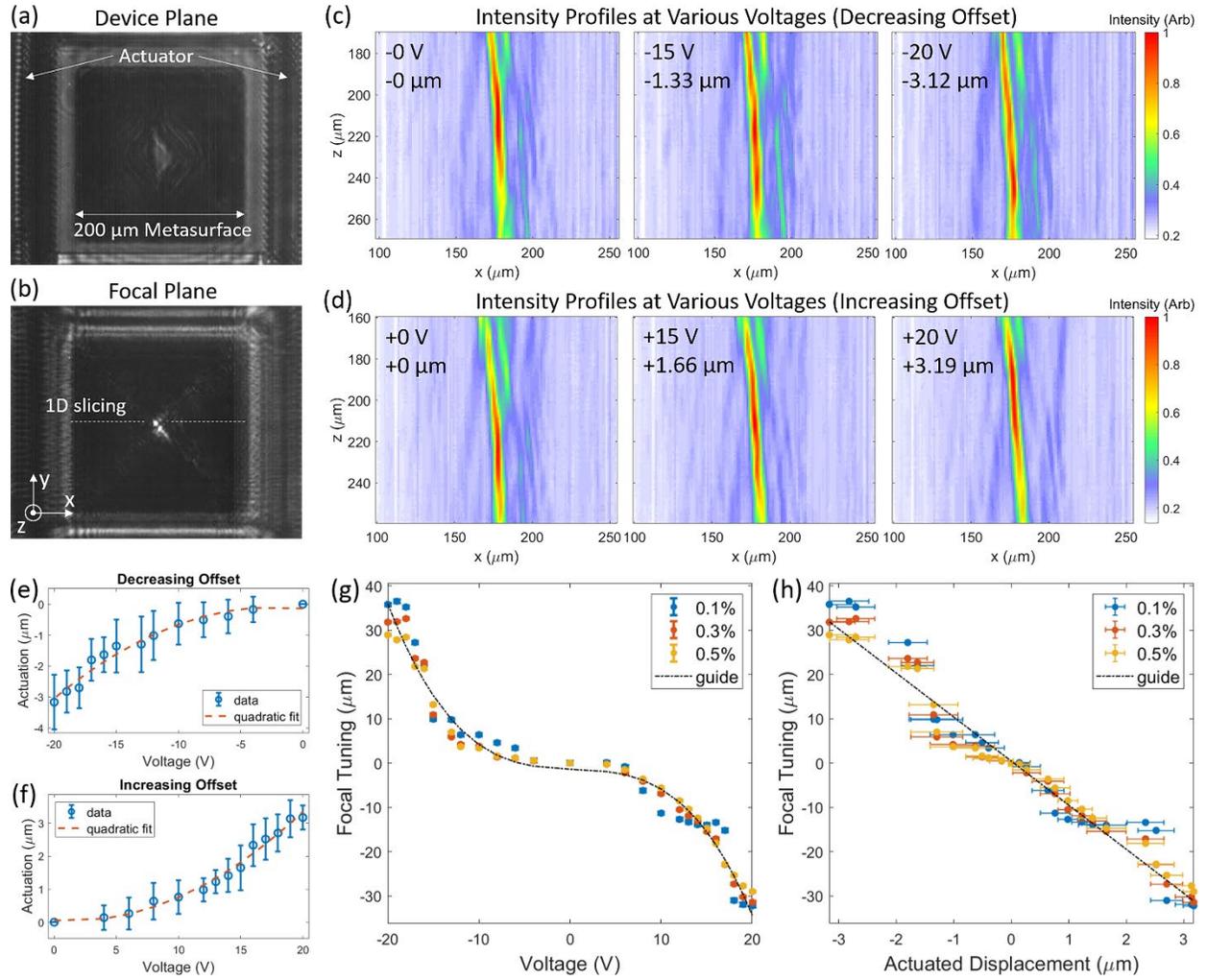

*Figure 4.* IR camera screenshots at (a) the device plane and (b) the focus plane of an Alvarez metalens illuminated with a 1550 nm SLD source. The shift of focal intensity profiles along the optical axis with (c) decreasing and (d) increasing center-to-center offsets between the Alvarez metasurfaces at various actuation voltages and displacements. (e, f) Actuated displacement as a function of the applied voltage with the negative (positive) actuation corresponding to a decreasing (increasing) offset. Tuning in focal location (g) as a function of the applied voltage and (h) as a function of the actuated displacement with the focus size defined by various intensity thresholds, giving a focal tuning range 10 times larger than the actuated Alvarez displacement. The black dash-dotted lines are provided as guides for the eye.

**Discussion**

Although in the work reported here, we have mainly focused on a MEMS-actuated Alvarez metalens with a 200 μm square aperture, Alvarez metalenses with larger apertures have been demonstrated with manual actuation[1,2]. Additionally, the Alvarez tuning principle is capable of continuously producing further focal tuning with an increasing lateral displacement between the two optical elements while maintaining high tuning efficiency. With the successful demonstration of the current MEMS Alvarez metalens, a larger lateral actuation can be achieved by reconfiguring the electrostatic MEMS actuator design or adjusting the actuation voltage below the pull-in threshold. The focusing behavior is significantly affected by the axial separation between the two Alvarez metasurfaces created during bonding. Uneven resin curing in this process can also lead to wedge errors that will shift or distort the focal spot. To mitigate these shortcomings, a thinner spacer layer deposited with high uniformity is desirable, and Alvarez metasurfaces with larger apertures can be implemented to improve performance tolerance to fabrication discrepancies.

While this work has demonstrated a MEMS-actuated metasurface Alvarez lens operating in the IR spectrum, other studies have reported a metasurface-based Alvarez lens operating in visible wavelengths when manually displaced on a fine translation stage[1]. While our current Alvarez metasurfaces are fabricated with silicon nitride nanoposts on a silicon membrane or substrate, the silicon underneath the Alvarez metasurfaces can be replaced with silicon oxide or quartz to create a clear optical path to realize visible wavelength operation.

**Conclusion**

We have designed, fabricated, and characterized a MEMS-actuated metasurface lens with a tunable focus that can be modulated based on the Alvarez tuning principle. With a voltage application below 20 V, an actuated displacement of 6.3 μm has given rise to focal length modulation over a range of 68 μm, which corresponds to a scaling over 10 times between the actuation input and focal tuning output. The MEMS electrostatic actuation using linear comb drives provides high controllability following the quadratic relation between the actuated displacement and actuation voltage. Replacing bulky freeform optics with flat, wavelength-scale thick metasurfaces together with the MEMS-driven Alvarez structure will allow the ultimate miniaturization of optical devices with a tunable focal length, making them more mechanically and thermally robust, with advantages such as fast tuning, compact size, simple packaging, and low energy consumption.

Such MEMS-actuated Alvarez metalenses with a tunable focus in IR or visible frequencies will have great potential in 3D imaging and depth-sensing applications, such as realizing mixed reality displays[25-27] and ultracompact endoscopes[19,28]. Additionally, by using a pair of quartic metasurfaces and computational imaging, the color information can be incorporated in a similar tunable geometry[29]. The full compatibility of the metalenses with well-established semiconductor microfabrication technologies allows them to be mass-produced with high reproducibility.

**Materials and Methods**

Metasurface Design

Silicon nitride nanoposts (see Figure S1a of the Supplementary Information) have been used as fundamental scattering elements because of their low absorption losses at both infrared and visible wavelengths due to their wide bandgap while exhibiting similar performance to other material platforms[24]. The transmission coefficients (see Figure S1b of the Supplementary Information) of the nanoposts for a fixed lattice constant of 1.3 μm and a post thickness of 2 μm on a silicon substrate were simulated as a function of the grating duty cycle using rigorous coupled-wave analysis (RCWA)[30]. The calculated phase spanned from 0 to $2\pi$ while maintaining a near-unity amplitude. The phase profile was then quantized into six linear steps from 0 to $2\pi$, giving the corresponding six cylindrical post diameters, which were achievable with stepper lithography, to construct the Alvarez cubic phase profiles for the two complementary metasurface optical elements presented in this work.

The metasurface singlet lens employed for motion tracking was designed with a similar procedure, using the same nanoposts with chosen diameters to map a quadratic phase profile for a converging singlet lens instead.

Device Fabrication

First, the two complementary Alvarez metasurfaces were fabricated in two parts: one actuatable metasurface half-plate carried by an electrostatic MEMS platform and one stationary metasurface half-plate on a silicon substrate. Then, the half-plates were aligned and bonded together to establish the Alvarez system. A schematic of the fabrication flow can be found in Figure S2 of the Supplementary Information.

The first Alvarez metasurface was fabricated on a silicon-on-insulator (SOI) wafer with a silicon device layer of 5-6 μm and a buried-oxide (BOX) layer of 2-4 μm. On top of the device layer, 2 μm of silicon nitride was deposited using plasma-enhanced chemical vapor deposition (PECVD). The first metasurface was patterned into silicon nitride with stepper lithography and inductively coupled plasma (ICP) etching. The MEMS electrostatic actuator to carry the metasurface was patterned into the silicon device layer and was etched by deep reactive ion etching (DRIE). Electrical contact pads and chip-level alignment marks were created at the designated locations by electron beam evaporation and lift-off of a gold/nickel/chromium (100 nm / 100 nm / 10 nm) stack. DRIE holes were etched from the backside of the wafer to remove the silicon substrate under the central metasurface platform, aiding the final hydrogen fluoride (HF) vapor etch to release the MEMS actuator.

The second Alvarez metasurface was fabricated on a double-side polished (DSP) silicon wafer. A 2 μm layer of PECVD silicon nitride was deposited and patterned with the second metasurface using stepper lithography and ICP etching. Chip-level alignment marks were etched through the silicon substrate with DRIE.

Wafers carrying the two metasurfaces were diced into individual chips. The chips with mobile metasurfaces were released with vapor HF etch. Then, a chip with the actuator carrying the first mobile metasurface was placed in parallel and was aligned face-to-face with a chip carrying the second stationary metasurface on a custom-built piezo-positioning stage, using 50 μm thick Kapton tape pieces to create a gap between the chips as the electrical insulation and to prevent accidental scratching between the metasurfaces. Once aligned, the chips were bonded by applying ultraviolet (UV)-sensitive resin along the stack edges to complete the assembly of the MEMS metasurface Alvarez lens. In the potential mass production of such Alvarez metalenses, the alignment and bonding of the metasurface pairs could be done on the wafer level by a commercial wafer bonding system with a deposited spacer layer before being diced into individual device chips.

Experimental Setup

Both tuning experiments of the MEMS metasurface singlet lens and the MEMS metasurface Alvarez lens used the same fundamental setup (see Figure S3 of the Supplementary Information).

The actuator chip was attached to a glass microscope slide using transparent UV resin. The glass slide was then mounted vertically onto a 3-axis translation stage that was manually moved by micropositioners for the initial alignment of the metalens to the overall setup. A source meter (Keithley 2450) grounded the chip substrates and applied the required voltages across the comb drives of the test devices for electrostatic actuation via probes to the respective contact pads. An SLD source of 1550 nm (Thorlabs S5FC1005P) was aligned with the optical axis of the metalens, illuminating it from one side. The incident light interacted with the Alvarez metasurfaces and converged to a focus along the optical axis. The light spreading from the focus was then collected and converted to parallel rays by the infinity-corrected objective (Mitutoyo M Plan Apo 20× Objective, NA = 0.42, f = 20 mm) on the other side of the metalens. The rays were then converged by a tube lens to be captured by the IR camera (Xenics Bobcat) that exported the live image to a control computer, which stored the data as sequential screenshots or video clips.

The objective, tube lens, and IR camera were mounted at fixed distances on a motorized linear stage (Newport ILS100CC), which was controlled by a motion controller (Newport ESP301) to change the distance between the optics and the metalens by steps as small as 1 μm. A nonzero voltage applied across the comb drive of the MEMS actuator laterally displaced the mobile metasurface, leading to focal tuning along the optical axis. The linear motorized stage was then used to scan along the optical axis to investigate the focal position as a function of the applied voltage.

**Acknowledgments**

This work was supported by funding provided by TunOptix Inc. Part of this work was conducted at the Washington Nanofabrication Facility (WNF)/Molecular Analysis Facility (MAF), a National Nanotechnology Coordinated Infrastructure (NNCI) site at the University of Washington, which was supported in part by funds from the National Science Foundation (awards NNCI-1542101, 1337840 and 0335765), the National Institutes of Health, the Molecular Engineering & Sciences Institute, the Clean Energy Institute, the Washington Research Foundation, the M. J. Murdock Charitable Trust, Altatech, ClassOne Technology, GCE Market, Google, and SPTS. The experimental setup and supporting data acquisition were partially assisted by James Whitehead and Luocheng Huang from the Nano Optoelectronic Integrated System Engineering (NOISE) Lab at the University of Washington.

**Conflict of Interest**

The authors declare no conflicts of interest.

**Author Contributions**

Zheyi Han designed, fabricated, and characterized the integrated miniature MEMS-actuated focal-tunable lens incorporating the Alvarez metasurfaces designed by Shane Colburn. Dr. Karl F. Böhringer and Dr. Arka Majumdar supervised the MEMS integration and Alvarez metasurface design, respectively, as the principal investigators.